\documentclass[sigconf,review=false]{acmart}

\usepackage{booktabs} 

\usepackage[english]{babel}
\usepackage{moresize}
\usepackage{amsmath}
\usepackage{algorithmic}
\usepackage{balance}
\usepackage{comment}
\usepackage{paralist}
\usepackage{bm}
\usepackage{pgfplots}
\usetikzlibrary{pgfplots.dateplot}

\usepackage{flushend}
\usepackage[english]{babel}
\usepackage[latin1]{inputenc}
\usepackage{mathrsfs}
\usepackage{graphicx}

\usepackage{amssymb}
\usepackage{amsfonts}
\usepackage{url}
\usepackage{longtable}
\usepackage{rotating}
\usepackage{multirow}
\usepackage{mathrsfs}
\usepackage{subfigure}
\usepackage{enumitem}
\usepackage[linesnumbered,algoruled,boxed,lined]{algorithm2e}
\usepackage{adjustbox}
\usepackage{hyperref}

\usepackage{pgfplots}
\usetikzlibrary{pgfplots.dateplot}

\usepackage{filecontents}
\definecolor{tblue}{RGB}{31,119,180}
\definecolor{torange}{RGB}{255,127,14}
\definecolor{tgreen}{RGB}{44,160,44}
\definecolor{tred}{RGB}{214,39,40}
\definecolor{tpurple}{RGB}{148,103,189}

\newcommand{\hide}[1]{} 

\newcommand{\ie}{\textit{i}.\textit{e}.}
\newcommand{\eg}{\textit{e}.\textit{g}.} 
\newcommand{\wrt}{\textit{w}.\textit{r}.\textit{t}}

\setcopyright{acmcopyright}

\def\model{SHT}

\keywords{Self-Supervised Learning, Graph Neural Networks, Hypergraph Representation, Recommender System}

\copyrightyear{2022}
\acmYear{2022}
\setcopyright{acmcopyright}\acmConference[KDD'22]{Proceedings of the 28th ACM SIGKDD Conference on Knowledge Discovery and Data Mining}{August 14--18, 2022}{Washington, DC, USA}
\acmBooktitle{Proceedings of the 28th ACM SIGKDD Conference on Knowledge Discovery and Data Mining (KDD'22), August 14--18, 2022, Washington, DC, USA}
\acmPrice{15.00}
\acmDOI{10.1145/3534678.3539473}
\acmISBN{978-1-4503-9385-0/22/08}


\begin{document}
\fancyhead{}

\title{Self-Supervised Hypergraph Transformer for \\ Recommender Systems}

\author{Lianghao Xia}
\affiliation{
  \institution{University of Hong Kong}
  \city{Hong Kong}
  \country{China}
}
\email{aka\_xia@foxmail.com}

\author{Chao Huang}
\authornote{Chao Huang is the corresponding author.}
\affiliation{
  \institution{University of Hong Kong}
  \city{Hong Kong}
  \country{China}
}
\email{chaohuang75@gmail.com}

\author{Chuxu Zhang}
\affiliation{
  \institution{Brandeis University}
  \city{Waltham}
  \country{USA}
}
\email{chuxuzhang@brandeis.edu}

\begin{abstract}
Graph Neural Networks (GNNs) have been shown as promising solutions for collaborative filtering (CF) with the modeling of user-item interaction graphs. The key idea of existing GNN-based recommender systems is to recursively perform the message passing along the user-item interaction edge for refining the encoded embeddings. Despite their effectiveness, however, most of the current recommendation models rely on sufficient and high-quality training data, such that the learned representations can well capture accurate user preference. User behavior data in many practical recommendation scenarios is often noisy and exhibits skewed distribution, which may result in suboptimal representation performance in GNN-based models. In this paper, we propose \model, a novel \underline{S}elf-Supervised \underline{H}ypergraph \underline{T}ransformer framework (\model) which augments user representations by exploring the global collaborative relationships in an explicit way. Specifically, we first empower the graph neural CF paradigm to maintain global collaborative effects among users and items with a hypergraph transformer network. With the distilled global context, a cross-view generative self-supervised learning component is proposed for data augmentation over the user-item interaction graph, so as to enhance the robustness of recommender systems. Extensive experiments demonstrate that \model\ can significantly improve the performance over various state-of-the-art baselines. Further ablation studies show the superior representation ability of our \model\ recommendation framework in alleviating the data sparsity and noise issues. The source code and evaluation datasets are available at: https://github.com/akaxlh/SHT.
\end{abstract}

\maketitle

\section{Introduction}
\label{sec:intro}

Recommender systems have become increasingly important to alleviate the information overload for users in a variety of web applications, such as e-commerce systems~\cite{guo2020debiasing}, streaming video sites~\cite{liu2021concept} and location-based lifestyle apps~\cite{chen2021curriculum}. To accurately infer the user preference, encoding user and item informative representations is the core part of effective collaborative filtering (CF) paradigms based on the observed user-item interactions~\cite{he2017neural,rendle2020neural,huang2021recent}.

Earlier CF models project interaction data into latent user and item embeddings using matrix factorization (MF)~\cite{koren2009matrix}. Due to the strong representation ability of deep learning, various neural network CF models have been developed to project users and items into latent low-dimensional representations, such as autoencoder~\cite{liang2018variational} and attention mechanism~\cite{chen2017attentive}. Recent years have witnessed the development of graph neural networks (GNNs) for modeling graph-structural data~\cite{wang2019heterogeneous,wu2019simplifying}. One promising direction is to perform the information propagation along the user-item interactions to refine user embeddings based on the recursive aggregation schema. For example, upon the graph convolutional network, PinSage~\cite{ying2018graph} and NGCF~\cite{wang2019neural} attempt to aggregate neighboring information by capturing the graph-based CF signals for recommendation. To simplify the graph-based message passing, LightGCN~\cite{he2020lightgcn} omits the burdensome non-linear transformer during the embedding propagation and improve the recommendation performance. To further enhance the graph-based user-item interaction modeling, some follow-up studies propose to learn intent-aware representations with disentangled graph neural frameworks (\eg, DisenHAN~\cite{wang2020disenhan}), differentiate behavior-aware embeddings of users with multi-relational graph neural models (\eg, MB-GMN~\cite{xia2021graph}).

Despite the effectiveness of the above graph-based CF models by providing state-of-the-art recommendation performance, several key challenges have not been well addressed in existing methods. \emph{First}, data noise is ubiquitous in many recommendation scenarios due to a variety of factors. For example, users may click their uninterested products due to the over-recommendation of popular items~\cite{zhang2021causal}. In such cases, the user-item interaction graph may contain `` interest-irrelevant'' connections. Directly aggregating information from all interaction edges will impair the accurate user representation. Worse still, the embedding propagation among multi-hop adjacent vertices (user or item) will amplify the noise effects, which misleads the encoding of underlying user interest in GNN-based recommender systems. \emph{Second}, data sparsity and skewed distribution issue still stand in the way of effective user-item interaction modeling, leading to most existing graph-based CF models being biased towards popular items~\cite{zhang2021model,krishnan2018adversarial}. Hence, the recommendation performance of current approaches severely
drops with the user data scarcity problem, as the high-quality training signals could be small. While there exist a handful of recently developed recommendation methods (SGL~\cite{wu2021self} and SLRec~\cite{yao2021self}) leveraging self-supervised learning to improve user representations, these methods mainly generate the additional supervision information with probability-based randomly mask operations, which might keep some noisy interaction and dropout some important training signals during the data augmentation process. \\\vspace{-0.12in}

\noindent \textbf{Contribution}. In light of the aforementioned challenges, this work proposes a \underline{S}elf-Supervised \underline{H}ypergraph \underline{T}ransformer (\model) to enhance the robustness and generalization performance of graph-based CF paradigms for recommendation. Specifically, we integrate the hypergraph neural network with the topology-aware Transformer, to empower our \model\ to maintain the global cross-user collaborative relations. Upon the local graph convolutional network, we first encode the topology-aware user embeddings and inject them into Transformer architecture for hypergraph-guided message passing within the entire user/item representation space. \\\vspace{-0.12in}

In addition, we unify the modeling of local collaborative relation encoder with the global hypergraph dependency learning under a generative self-supervised learning framework. Our proposed new self-supervised recommender system distills the auxiliary supervision signals for data augmentation through a graph topological denoising scheme. A graph-based meta transformation layer is introduced to project hyergraph-based global-level representations into the graph-based local-level interaction modeling for user and item dimensions. Our new proposed \model\ is a model-agnostic method and serve as a plug-in learning component in existing graph-based recommender systems. Specifically, \model\ enables the cooperation of the local-level and global-level collaborative relations, to facilitate the graph-based CF models to learn high-quality user embeddings from noisy and sparse user interaction data.

The key contributions of this work are summarized as follows:
\begin{itemize}[leftmargin=*]

\item In this work, we propose a new self-supervised recommendation model--\model\ to enhance the robustness of graph collaborative filtering paradigms, by integrating the hypergraph neural network with the topology-aware Transformer.\\\vspace{-0.1in}

\item In the proposed \model\ method, the designed hypergraph learning component encodes the global collaborative effects within the entire user representation space, via a learnable multi-channel hyperedge-guided message passing schema. Furthermore, the local and global learning views for collaborative relations are integrated with the cooperative supervision for interaction graph topological denoising and auxiliary knowledge distillation. \\\vspace{-0.1in}

\item Extensive experiments demonstrate that our proposed \model\ framework achieves significant performance improvement over 15 different types of recommendation baselines. Additionally, we conduct empirical analysis to show the rationality of our model design with the ablation studies.
\end{itemize}
\section{Preliminaries and Related Work}
\label{sec:relate}

\noindent \textbf{Recap Graph Collaborative Filtering Paradigm}. To enhance the Collaborative Filtering with the multi-order connectivity information, one prominent line of recommender systems generates graph structures for user-item interactions. Suppose our recommendation scenario involves $I$ users and $J$ items with the user set $\mathcal{U}=\{u_1,...u_I\}$ and item set $\mathcal{V}=\{v_1,...v_J\}$. Edges in the user-item interaction graph $\mathcal{G}$ are constructed if user $u_i$ has adopted item $v_j$. Upon the constructed interaction graph structures, the core component of graph-based CF paradigm lies in the information aggregation function--gathering the feature embeddings of neighboring users/items via different aggregators, \eg, mean or sum.\\\vspace{-0.12in}

\noindent \textbf{Recommendation with  Graph Neural Networks}. Recent studies have attempted to design various graph neural architectures to model the user-item interaction graphs through embedding propagation. For example, PinSage~\cite{ying2018graph} and NGCF~\cite{wang2019neural} are built upon the graph convolutional network over the spectral domain. Later on, LightGCN~\cite{he2020lightgcn} proposes to simplify the heavy non-linear transformation and utilizes the sum-based pooling over neighboring representations. Upon the GCN-based message passing schema, each user and item is encoded into the transformed embeddings with the preservation of multi-hop connections. To further improve the user representation, some recent studies attempt to design disentangled graph neural architecture for user-item interaction modeling, such as DGCF~\cite{wang2020disentangled} and DisenHAN~\cite{wang2020disenhan}. Several multi-relational GNNs are proposed to enhance recommender systems with multi-behavior modeling, including KHGT~\cite{xia2021knowledge} and HMG-CR~\cite{yang2021hyper}. However, most of existing graph neural CF models are intrinsic designed to merely rely on the observation interaction lables for model training, which makes them incapable of effectively modeling interaction graph with sparse and noisy supervision signals. To overcome these challenges, this work proposes a self-supervised hypergraph transformer architecture to generate informative knowledge through the effective interaction between local and global collaborative views. \\\vspace{-0.12in}

\noindent \textbf{Hypergraph-based Recommender Systems}. There exist some recently developed models constructing hypergraph connections to improve the relation learning for recommendation~\cite{wang2020next,ji2020dual,yu2021self}. For example, HyRec~\cite{wang2020next} regards users as hyperedges to aggregate information from the interacted items. MHCN~\cite{yu2021self} constructs multi-channel hypergraphs to model high-order relationships among users. Furthermore, DHCF~\cite{ji2020dual} is a hypergraph collaborative filtering model to learn the hybrid high-order correlations. Different from these work for generating hypergraph structures with manually design, this work automates the hypergraph structure learning process with the modeling of global collaborative relation. \\\vspace{-0.12in}

\noindent \textbf{Self-Supervised Graph Learning}.
To improve the embedding quality of supervised learning, self-supervised learning (SSL) has become a promising solution with auxiliary training signals~\cite{liu2021self}, such as augmented image data~\cite{kang2020contragan}, pretext sequence tasks for language data~\cite{vulic2021lexfit}, knowledge graph augmentation~\cite{yang2022knowledge}. Recently, self-supervised learning has also attracted much attention on graph representation~\cite{hwang2020self}. For example, DGI~\cite{velickovic2019deep} and GMI~\cite{peng2020graph} perform the generative self-supervised learning over the GNN framework with auxiliary tasks. Inspired by the graph self-supervised learning, SGL~\cite{wu2021self} produces state-of-the-art performance by generating contrastive views with randomly node and edge dropout operations. Following this research line, HCCF~\cite{xia2022hypergraph} leverages the hypergraph to generate contrastive signals to improve the graph-based recommender system. Different from them, this work enhances the graph-based collaborative filtering paradigm with a generative self-supervised learning framework.

\section{Methodology}
\label{sec:solution}

\begin{figure*}
    \centering
    \includegraphics[width=1.0\textwidth]{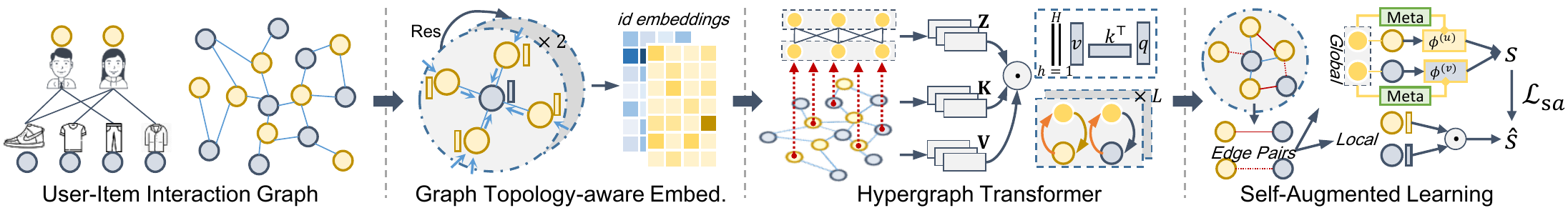}
    \vspace{-0.2in}
    \caption{Overall framework of the proposed \model\ model.}
    \vspace{-0.1in}
    \label{fig:framework}
\end{figure*}

In this section, we present the proposed \model\ framework and show the overall model architecture in Figure~\ref{fig:framework}. \model\ embeds local structure information into latent node representations, and conduct global relation learning with the local-aware hypergraph transformer. To train the proposed model, we augment the regular parameter learning with the local-global cross-view self-augmentation.

\subsection{Local Graph Structure Learning}
To begin with, we embed users and items into a $d$-dimensional latent space to encode their interaction patterns. For user $u_i$ and item $v_j$, embedding vectors $\textbf{e}_i, \textbf{e}_j \in \mathbb{R}^d$ are generated, respectively. Also, we aggregate all the user and item embeddings to compose embedding matrices $\textbf{E}^{(u)} \in \mathbb{R}^{I\times d}, \textbf{E}^{(v)} \in \mathbb{R}^{J\times d}$, respectively. We may omit the superscript $(u)$ and $(v)$ for notation simplification when it is not important to differentiate the user and item index.

Inspired by recent success of graph convolutional networks~\cite{wu2019simplifying, he2020lightgcn} in capturing local graph structures, we propose to encode the neighboring sub-graph structure of each node into a graph topology-aware embedding, to inject the topology positional information into our graph transformer. Specifically, \model\ employs a two-layer light-weight graph convolutional network as follows:
\begin{align}
    \bar{\textbf{E}}^{(u)} = \text{GCN}^2(\textbf{E}^{(v)}, \mathcal{G})= \bar{\mathcal{A}} \cdot \bar{\mathcal{A}}^\top \textbf{E}^{(u)} + \bar{\mathcal{A}} \cdot \textbf{E}^{(v)}
\end{align}
where $\bar{\textbf{E}}^{(u)}\in\mathbb{R}^{I\times d}$ denotes the topology-aware embeddings for users. $\text{GCN}^2(\cdot)$ denotes two layers of message passing. $\bar{\mathcal{A}}\in\mathbb{R}^{I\times J}$ refers to the normalized adjacent matrix of graph $\mathcal{G}$, which is calculated by $\bar{\mathcal{A}}_{i,j}=\mathcal{A}_{i,j} / (\textbf{D}_i^{(u)1/2} \textbf{D}_j^{(v)1/2})$, where $\mathcal{A}$ is the original binary adjacent matrix. $\textbf{D}_i^{(u)}, \textbf{D}_j^{(v)}$ refer to the degree of $u_i$ and $v_j$ in graph $\mathcal{G}$, respectively. Note that \model\ considers neighboring nodes in different distance through residual connections. The topology-aware embeddings for items can be calculated analogously.

\subsection{Hypergraph Transformer for Global Relation Learning}
Though existing graph-based neural networks have shown their strength in learning interaction data~\cite{wang2019neural,he2020lightgcn,chen2020revisiting}, the inherent noise and skewed data distribution in recommendation scenario limit the performance of graph representation for user embeddings. To address this limitation, \model\ adopts a hypergraph transformer framework, which i) alleviates the noise issue by enhancing the user collaborative relation modeling with the adaptive hypergraph relation learning; ii) transfer knowledge from dense user/item nodes to sparse ones.
Concretely, \model\ is configured with a Transformer-like attention mechanism for structure learning. The encoded graph topology-aware embeddings are injected into the node representations to preserve the graph locality and topological positions. Meanwhile, the multi-channel attention~\cite{sun2019bert4rec} further benefits our structure learning in \model.

In particular, \model\ generates input embedding vectors for $u_i$ and $v_j$ by combining the id-corresponding embeddings ($\textbf{e}_i, \textbf{e}_j$) together with the topology-aware embeddings ( vectors $\bar{\textbf{e}}_i, \bar{\textbf{e}}_j$ from embedding tables $\bar{\textbf{E}}^{(u)}$ and $\bar{\textbf{E}}^{(v)}$) as follows:
\begin{align}
    \tilde{\textbf{e}}_i = \textbf{e}_i + \bar{\textbf{e}}_i;~~~~~~
    \tilde{\textbf{e}}_j = \textbf{e}_j + \bar{\textbf{e}}_j
\end{align}
Then, \model\ conducts hypergraph-based information propagation as well as hypergraph structure learning using $\tilde{\textbf{e}}_i, \tilde{\textbf{e}}_j$ as input. We utilize $K$ hyperedges to distill the collaborative relations from the global perspective. Node embeddings are propagated to each other using hyperedges as intermediate hubs, where the connections between nodes and hyperedges are optimized to reflect the implicit dependencies among nodes.

\subsubsection{\bf Node-to-Hyperedge Propagation}
Without loss of generality, we mainly discuss the information propagation between user nodes and user-side hyperedges for simplicity. The same process is applied for item nodes analogously. The propagation from user nodes to user-side hyperedges can be formally presented as follows:
\begin{align}
    \tilde{\textbf{z}}_k = \mathop{\Bigm|\Bigm|}\limits_{h=1}^H \bar{\textbf{z}}_{k,h};~~~~
    \bar{\textbf{z}}_{k,h} = \sum_{i=1}^I \textbf{v}_{i,h} \textbf{k}_{i,h}^\top \textbf{q}_{k,h}
\end{align}
where $\tilde{\textbf{z}}_k\in\mathbf{R}^d$ denotes the embedding for the $k$-th hyperedge. It is calculated by concatenating the $H$ head-specific hyperedge embeddings $\bar{\textbf{z}}_{k,h} \in \mathbb{R}^{d/H}$. $\textbf{q}_{k,h}, \textbf{k}_{i,h}, \textbf{v}_{i,h}\in\mathbb{R}^{d/H}$ are the query, key and value vectors in the attention mechanism which will be elaborated later. Here, we calculate the edge weight between hyperedge $k$ and user $u_i$ through a linear dot-product $\textbf{k}_{i,h}^\top \textbf{q}_{k,h}$, which reduces the complexity from $O(K\times I\times d / H)$ to $O((I+K)\times d^2/ {H^2})$ by avoiding directly calculating the node-hyperedge connections (\ie~$\textbf{k}_{i,h}^\top \textbf{q}_{k,h}$), but the key-value dot-product first (\ie~$\sum_{i=1}^I \textbf{v}_{i,h} \textbf{k}_{i,h}^\top$).

In details, the multi-head query, key and value vectors are calculated through linear transformations and slicing. The $h$-head-specific embeddings are calculated by:
\begin{align}
    \label{eq:qkv}
    \textbf{q}_{k,h}=\textbf{Z}_{k, p_{h-1}: p_h};~~~
    \textbf{k}_{i,h}=\textbf{K}_{p_{h-1}: p_h, :} \tilde{\textbf{e}}_i;~~~
    \textbf{v}_{i,h}=\textbf{V}_{p_{h-1}: p_h, :} \tilde{\textbf{e}}_i
\end{align}
where $\textbf{q}_{k,h}\in\mathbb{R}^{d/H}$ denotes the $h$-head-specific query embedding for the $k$-th hyperedge, $\textbf{k}_{i,h}, \textbf{v}_{i,h} \in\mathbb{R}^{d/H}$ denotes the $h$-head-specific key and value embedding for user $u_i$. $\textbf{Z}\in\mathbb{R}^{K\times d}$ represents the embedding matrix for the $K$ hyperedges. $\textbf{K}, \textbf{V} \in\mathbb{R}^{d\times d}$ represents the key and the value transformation of all the $H$ heads, respectively.
$p_{h-1} = \frac{(h-1)d}{H}$ and $p_h = \frac{hd}{H}$ denote the start and the end indices of the $h$-th slice.

To further excavate the complex non-linear feature interactions among the hyperedges, \model\ is augmented with two-layer hierarchical hypergraph neural networks for both user side and item side. Specifically, the final hyperedge embeddings are calculated by:
\begin{align}
    \hat{\textbf{Z}} = \text{HHGN}^2(\tilde{\textbf{Z}});~~~~
    \text{HHGN}(\textbf{X}) = \sigma(\mathcal{H}\cdot \textbf{X} + \textbf{X})
\end{align}
where $\hat{\textbf{Z}}, \tilde{\textbf{Z}} \in\mathbb{R}^{K\times d}$ represent the embedding tables for the final and the original hyperedge embeddings, consisting of hyperedge-specific embedding vectors $\hat{\textbf{z}}, \tilde{\textbf{z}}\in\mathbb{R}^d$, respectively.
$\text{HHGN}^2(\cdot)$ denotes applying the hierarchical hypergraph network (HHGN) twice. HHGN is configured with a learnable parametric matrix $\mathcal{H}\in\mathbb{R}^{K \times K}$, which characterizes the hyperedge-wise relations. An activation function $\sigma(\cdot)$ is introduced for non-linear relation modeling. Additionally, we utilize a residual connection to facilitate gradient propagation in our hypergraph neural structures.

\subsubsection{\bf Hyperedge-to-Node Propagation}
With the final hyperedge embeddings $\hat{\textbf{Z}}$, we propagate the information from hyperedges to user/item nodes through a similar but reverse process:
\begin{align}
    \tilde{\textbf{e}}_i' = \mathop{\Bigm|\Bigm|}\limits_{h=1}^H \bar{\textbf{e}}_{i,h}';~~~~
    \bar{\textbf{e}}_{i,h}' = \sum_{k=1}^K \textbf{v}_{k,h}' {\textbf{k}'}_{k,h}^{\top} \textbf{q}_{i,h}'
\end{align}
where $\tilde{\textbf{e}}_i'\in\mathbb{R}^d$ denotes the new embedding for user $u_i$ refined by the hypergraph neural network. $\bar{\textbf{e}}_{i,h}'\in\mathbb{R}^{d/H}$ denotes the node embedding calculated by the $h$-th attention head for $u_i$. $\textbf{q}_{i,h}', \textbf{k}_{k,h}', \textbf{v}_{k,h}' \in\mathbb{R}^{d/H}$ represent the query, key and value vectors for user $u_i$ and hyperedge $k$. The attention calculation in this hyperedge-to-node propagation process shares most parameters with the aforementioned node-to-hyperedge propagation. The former query serves as key, and the former key serves as query here. The value calculation applies the same value transformation for the hyperedge embedding. The calculation process can be formally stated as:
\begin{align}
    \textbf{q}_{i,h}' = \textbf{k}_{i,h};~~~~
    \textbf{k}_{k,h}' = \textbf{q}_{k,h};~~~~
    \textbf{v}_{k,h}' = \textbf{V}_{p_{h-1}:p_h,:} \hat{\textbf{z}}_k
\end{align}

\subsubsection{\bf Iterative Hypergraph Propagation}
Based on the prominent node-wise relations captured by the learned hypergraph structures, we propose to further propagate the encoded global collaborative relations via stacking multiple hypergraph transformer layers. In this way, the long-range user/item dependencies can be characterized by our \model\ framework through the iterative hypergraph propagation. In form, taking the embedding tables $\tilde{\textbf{E}}_{l-1}$ in the $(l-1)$-th iteration as input, \model\ recursively applies the hypergraph encoding (denoted by $\text{HyperTrans}(\cdot)$) and obtains the final node embeddings $\hat{\textbf{E}}\in\mathbb{R}^{I\times d}$ or $\mathbb{R}^{J\times d}$ as follows:
\begin{align}
    \tilde{\textbf{E}}_l = \text{HyperTrans}(\tilde{\textbf{E}}_{l-1});~~~~
    \hat{\textbf{E}} = \sum_{l=1}^L \tilde{\textbf{E}}_l
\end{align}
\noindent where the layer-specific embeddings are combined through element-wise summation. The iterative hypergraph propagation is identical for the user nodes and item nodes. Finally, \model\ makes predictions through dot product as $p_{i,j} = \hat{\textbf{e}}_i^{(u)\top}\hat{\textbf{e}}_j^{(v)}$, where $p_{i,j}$ is the forecasting score denoting the probability of $u_i$ interacting with $v_j$.

\subsection{Local-Global Self-Augmented Learning}
\begin{figure}[t]
    \centering
    \includegraphics[width=0.9\columnwidth]{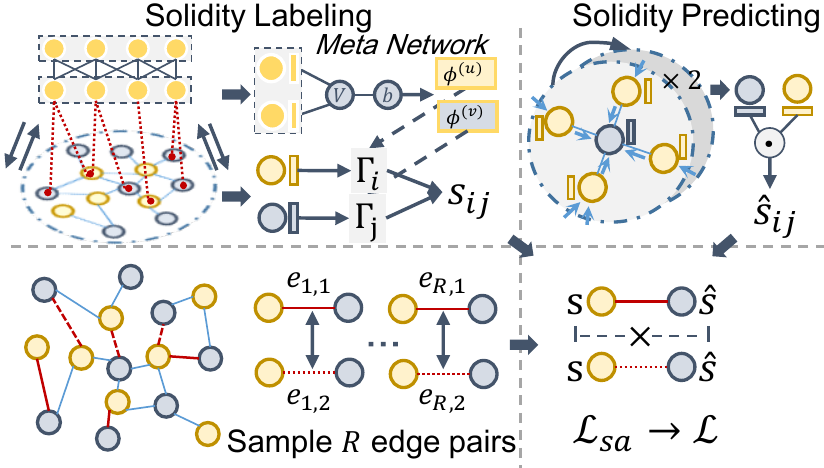}
    \vspace{-0.1in}
    \caption{Workflow of the self-augmented learning.}
    \vspace{-0.25in}
    \label{fig:framework_sal}
\end{figure}

The foregoing hypergraph transformer addresses the data sparsity problem through adaptive hypergraph message passing. However, the graph topology-aware embedding for local collaborative relation modeling may still be affected by the interaction data noise. To tackle this challenge, we propose to enhance the model training with self-augmented learning between the local topology-aware embedding and the global hypergraph learning. To be specific, the topology-aware embedding for local information extraction is augmented with an additional task to differentiate the solidity of sampled edges in the observed user-item interaction graph. Here, solidity refers to the probability of an edge not being noisy, and its label in the augmented task is calculated based on the learned hypergraph dependencies and representations. In this way, \model\ transfers knowledge from the high-level and denoised features in the hypergraph transformer, to the low-level and noisy topology-aware embeddings, which is expected to recalibrate the local graph structure and improve the model robustness. The workflow of our self-augmented module is illustrated in Fig~\ref{fig:framework_sal}.

\subsubsection{\bf Solidity Labeling with Meta Networks}
In our \model\ model, the learned hypergraph dependency representations can serve as useful knowledge to denoise the observed user-item interactions by associating each edge with a learned solidity score. Specifically, we reuse the key embeddings $\textbf{k}_{i,h}, \textbf{k}_{j,h}$ in Eq~\ref{eq:qkv} to represent user $u_i$ and item $v_j$ when estimating the solidity score for the edge $(u_i, v_j)$. This is because that the key vectors are generated for relation modeling and can be considered as helpful information source for interaction solidity estimation. Furthermore, we propose to also take the hyperedge embeddings $\textbf{Z}\in\mathbb{R}^{K\times d}$ in Eq~\ref{eq:qkv} into consideration, to introduce global characteristics into the solidity labeling.



Concretely, we first concatenate the multi-head key vectors and apply a simple perceptron to eliminate the gap between user/item-hyperedge relation learning and user-item relation learning. Formally, the updated user/item embeddings are calculated by:
\begin{align}
    \mathbf{\Gamma}_i=\phi^{(u)}\left(\mathop{\Bigm|\Bigm|}\limits_{h=1}^H \textbf{k}_{i,h}\right); ~~~~~~
    \mathbf{\Gamma}_j=\phi^{(v)}\left(\mathop{\Bigm|\Bigm|}\limits_{h=1}^H \textbf{k}_{j,h}\right)
\end{align}
\noindent where $\phi^{(u)}(\cdot), \phi^{(v)}(\cdot)$ are the user- and item-specific perceptrons for feature vector transformation, respectively. This projection is conducted with a meta network, using the user-side and the item-side hyperedge embeddings as input individually:
\begin{align}
    \phi(\textbf{x}; \textbf{Z}) = \sigma(\textbf{W} \textbf{x} + \textbf{b});~~
    \textbf{W} = \textbf{V}_1 \bar{\textbf{z}} + \textbf{W}_0;~~
    \textbf{b} = \textbf{V}_2 \bar{\textbf{z}} + \textbf{b}_0
\end{align}
\noindent where $\textbf{x}\in\mathbb{R}^d$ denotes the input user/item key embedding (\eg~$\mathbf{\Gamma}_i, \mathbf{\Gamma}_j$). $\phi(\cdot)$ being user-specific or item-specific depends on $\textbf{Z}$ being user-side or item-side hyperedge embedding table. $\textbf{W}\in\mathbb{R}^{d\times d}$ and $\textbf{b}\in\mathbb{R}^d$ are the parameters generated by the meta network according to the input $\textbf{Z}$. In this way, the parameters are generated based on the learned hyperedge embeddings, which encodes global features of user- or item-specific hypergraphs. $\bar{\textbf{z}}\in\mathbb{R}^{d}$ denotes the mean pooling of hyperedge embeddings (\ie~$\bar{\textbf{z}} = \sum_{k=1}^K \textbf{z}_k / K$). $\textbf{V}_1\in\mathbb{R}^{d\times d\times d}, \textbf{W}_0\in\mathbb{R}^{d\times d}, \textbf{V}_2\in\mathbb{R}^{d\times d}, \textbf{b}_0\in\mathbb{R}^{d}$ are the parameters of the meta network.

With the updated user/item embeddings $\mathbf{\Gamma}_i, \mathbf{\Gamma}_j$, \model\ then calculates the solidity labels for edge $(u_i, v_j)$ through a two-layer neural network as follows:
\begin{align}
    s_{i,j} = \text{sigm}(\textbf{d}^\top \cdot \sigma(\textbf{T} \cdot [\mathbf{\Gamma}_i; \mathbf{\Gamma}_j] + \mathbf{\Gamma}_i + \mathbf{\Gamma}_j + \textbf{c}))
\end{align}
\noindent where $s_{i,j}\in\mathbb{R}$ denotes the solidity score given by the hypergraph transformer. $\text{sigm}(\cdot)$ denotes the sigmoid function which limits the value range of $s_{i,j}$. $\textbf{d}\in\mathbb{R}^d, \textbf{T}\in\mathbb{R}^{d\times 2d}, \textbf{c}\in\mathbb{R}^d$ are the parametric matrices or vectors. $[\cdot;\cdot]$ denotes the vector concatenation.

\subsubsection{\bf Pair-wise Solidity Ranking}
To enhance the optimization of topological embeddings, \model\ employs an additional objective function to better estimate the edge solidity using the above $s_{i,j}$ as training labels. In particular, $R$ pairs of edges $\{(e_{1, 1}, e_{1,2})$,...,$(e_{R,1}, e_{R,2})\}$ from the observed edges in $\mathcal{G}$ are sampled, and \model\ gives predictions on the solidity using the topology-aware embeddings. The predictions are then updated by optimizing the loss below:
\begin{align}
    \label{eq:sa_loss}
    \mathcal{L}_{sa} = &\sum_{r=1}^R \text{max}(0, 1 - (\hat{s}_{u_{r,1}, v_{r,1}} - \hat{s}_{u_{r,2}, v_{r,2}}) (s_{u_{r,1}, v_{r,1}} - s_{u_{r, 2}, v_{r, 2}}));\nonumber\\
    &\hat{s}_{u_{r,1}, v_{r,1}} = \textbf{e}_{u_{{r,1}}}^\top \textbf{e}_{v_{{r, 1}}}; ~~~~~~
    \hat{s}_{u_{r,2}, v_{r,2}} = \textbf{e}_{u_{{r,2}}}^\top \textbf{e}_{v_{{r, 2}}}
\end{align}
\noindent where $\mathcal{L}_{sa}$ denotes the loss function for our self-augmented learning. $\hat{s}_{u_{r,1}, v_{r,1}}, \hat{s}_{u_{r,2}, v_{r,2}}$ denote the solidity scores predicted by the topology-aware embedding, while ${s}_{u_{r,1}, v_{r,1}}, {s}_{u_{r,2}, v_{r,2}}$ denote the edge solidity labels given by the hypergraph transformer. Here $u_{r,1}$ and $v_{r,1}$ represent the user and the item node of edge $e_{r,1}$, respectively.

In the above loss function, the label term $(s_{u_{r,1}, v_{r,1}} - s_{u_{r,2}, v_{r,2}})$ not only indicates the sign of the difference (\ie~which one of $e_{r,1}$ and $e_{r,2}$ is bigger), but also indicates how bigger the difference is. In this way, if the solidity labels for a pair of edges given by the hypergraph transformer are close to each other, then the gradients on the predicted solidity scores given by the topology-aware embedding will become smaller. In this way, \model\ is self-augmented with an adaptive ranking task, to further refine the low-level topology-aware embeddings using the high-level embeddings encoded from the hypergraph transformer.

\subsection{Model Learning}
We train our \model\ by optimizing the main task on implicit feedback together with the self-augmented ranking task. Specifically, $R'$ positive edges (observed in $\mathcal{G}$) and $R'$ negative edges (not observed in $\mathcal{G}$) are sampled $\{(e_{1, 1}, e_{1, 2}), (e_{2,1}, e_{2, 2})..., (e_{R',1}, e_{R',2})\}$, where $e_{r,1}$ and $e_{r,2}$ are individual positive and negative sample, respectively. The following pair-wise marginal objective function is applied:
\begin{align}
    \label{eq:loss}
    \mathcal{L} = \sum_{r=1}^{R'} \text{max}(0, 1-(p_{u_{r,1}, v_{r,1}} - p_{u_{r,2}, v_{r,2}})) + \lambda_1 \mathcal{L}_{\text{sa}} + \lambda_2 \|\mathbf{\Theta}\|_\text{F}^2
\end{align}
\noindent where $p_{u_{r,1}, v_{r,1}}$ and $p_{u_{r,2}, v_{r,2}}$ are prediction scores for edge $e_{r,1}$ and $e_{r,2}$, respectively. $\lambda_1$ and $\lambda_2$ are weights for different loss terms. $\|\mathbf{\Theta}\|_\text{F}^2$ denotes the $l_2$ regularization term for weight decay.

\subsubsection{\bf Complexity Analysis}
We compare our \model\ framework with several state-of-the-art approaches on collaborative filtering, including graph neural architectures (\eg~NGCF~\cite{wang2019neural}, LightGCN~\cite{he2020lightgcn}) and hypergraph neural networks(\eg~DHCF~\cite{ji2020dual}). As discussed before, our hypergraph transformer enables the complexity reduction from $O(K\times (I + J) \times d)$ to $O((I+J+K) \times d^2)$. As the typical value of the number of hyperedge $K$ is much smaller than the number of nodes $I$ and $J$, but larger than the embedding size $d$, the latter term is smaller and close to $O((I+J)\times d^2)$. In comparison, the complexity for a typical graph neural architecture is $O(M\times d + (I+J)\times d^2)$. So our hypergraph transformer network can achieve comparable efficiency as GNNs, such as graph convolutional networks in model inference. The existing hypergraph-based methods commonly pre-process high-order node relations to construct hypergraphs, which makes them usually more complex than the graph neural networks. In our \model, the self-augmented task with the loss $\mathcal{L}_\text{sa}$ has the same complexity with the original main task. 
\section{Evaluation}
\label{sec:eval}
To evaluate the effectiveness of our \model, our experiments are designed to answer the following research questions:
\begin{itemize}[leftmargin=*]
\item \textbf{RQ1}: How does our \model\ perform by comparing to strong baseline methods of different categories under different settings?
\item \textbf{RQ2}: How do the key components of \model\ (\eg, the hypergraph modeling, the transformer-like information propagation) contribute to the overall performance of \model\ on different datasets?
\item \textbf{RQ3}: How well can our \model\ handle noisy and sparse data, as compared to baseline methods?
\item \textbf{RQ4}: In real cases, can the designed the self-supervised learning mechanism in \model\ provide useful interpretations?
\end{itemize}

\begin{table}[]
    \centering
    \small
    \caption{Statistical information of the experimental datasets.}
    \vspace{-0.15in}
    \begin{tabular}{lcccc}
        \toprule
        Stat. & Yelp & Gowalla & Tmall\\
        \midrule
        \# Users & 29601 & 50821 & 47939\\
        \# Items & 24734& 24734 & 41390\\
        \# Interactions & 1517326 & 1069128 & 2357450\\
        Density & $2.1\times 10^{-3}$ & $4.0\times 10^{-4}$ & $1.2\times 10^{-3}$\\
        \bottomrule
    \end{tabular}
    \vspace{-0.1in}
    \label{tab:data}
\end{table}

\subsection{Experimental Settings}
\subsubsection{\bf Experimental Datasets}
The experiments are conducted on three datasets collected from real-world applications, \ie, Yelp, Gowalla and Tmall. The statistics of them are shown in Table~\ref{tab:data}.
\begin{itemize}[leftmargin=*]
\item \textbf{Yelp}: This commonly-used dataset contains user ratings on business venues collected from Yelp. Following other papers on implicit feedback~\cite{huang2021knowledge}, we treat users' rated venues as interacted items and treat unrated venues as non-interacted items.

\item \textbf{Gowalla}: It contains users' check-in records on geographical locations obtained from Gowalla. This evaluation dataset is generated from the period between 2016 and 2019.

\item \textbf{Tmall}: This E-commerce dataset is released by Tmall, containing users' behaviors for online shopping. We collect the page-view interactions during December in 2017.
\end{itemize}

\subsubsection{\bf Evaluation Protocols}
Following the recent collaborative filtering models~\cite{he2020lightgcn, wu2021self}, we split the datasets by 7:2:1 into training, validation and testing sets. We adopt all-rank evaluation protocol. When testing a user, the positive items in the test set and all the non-interacted items are tested and ranked together. We employ the commonly-used \textit{Recall@N} and \textit{Normalized Discounted Culmulative Gain (NDCG)@N} as evaluation metrics for recommendation performance evaluation~\cite{wang2019neural,ren2020sequential}. \textit{N} is set as 20 by default.

\subsubsection{\bf Compared Baseline Methods}
We evaluate our \model\ by comparing it with 15 baselines from different research lines for comprehensive evaluation.

\noindent \textbf{Traditional Factorization-based Technique.}
\begin{itemize}[leftmargin=*]

\item \textbf{BiasMF}~\cite{koren2009matrix}: This method augments matrix factorization with user and item bias vectors to enhance user-specific preferences.
\end{itemize}

\noindent \textbf{Neural Factorization Method.}\vspace{-0.05in}
\begin{itemize}[leftmargin=*]
\item \textbf{NCF}~\cite{he2017neural}: This method replaces the dot-product in conventional matrix factorization with multi-layer neural networks. Here, we adopt the NeuMF variant for comparison.
\end{itemize}

\noindent \textbf{Autoencoder-based Collaborative Filtering Approach.}
\begin{itemize}[leftmargin=*]
\item \textbf{AutoR}~\cite{sedhain2015autorec}: It improves user/item representations with a three-layer autoencoder trained under a behavior reconstruction task.
\end{itemize}

\noindent \textbf{Graph Neural Networks for Recommendation.}
\begin{itemize}[leftmargin=*]
\item \textbf{GCMC}~\cite{berg2017graph}: This is one of the pioneering work to apply graph convolutional networks (GCNs) to the matrix completion task.
\item \textbf{PinSage}~\cite{ying2018graph}: It applies random sampling in graph convolutional framework to study the collaborative filtering task
.
\item \textbf{NGCF}~\cite{wang2019neural}: This graph convolution-based approach additionally takes source-target representation interaction learning into consideration when designing its graph encoder.

\item \textbf{STGCN}~\cite{zhang2019star}: The model combines conventional graph convolutional encoders with graph autoencoders to improve the model robustness against sparse and cold-start samples.

\item \textbf{LightGCN}~\cite{he2020lightgcn}: This work conducts in-depth analysis to study the effectiveness of modules in standard GCN for collaborative data, and proposes a simplified GCN model for recommendation.

\item \textbf{GCCF}~\cite{chen2020revisiting}: This is another method which simplifies the GCNs by removing the non-linear transformation. In GCCF, the effectiveness of residual connections across graph iterations is validated.
\end{itemize}

\noindent \textbf{Disentangled Graph Model for Recommendation.}
\begin{itemize}[leftmargin=*]
\item \textbf{DGCF}~\cite{wang2020disentangled}: It disentangles user interactions into multiple latent intentions to model user preference in a fine-grained way.
\end{itemize}

\noindent \textbf{Hypergraph-based Neural Collaborative Filtering.}
\begin{itemize}[leftmargin=*]
\item \textbf{HyRec}~\cite{wang2020next}: This is a sequential collaborative model that learns item-wise high-order relations with hypergraphs.
\item \textbf{DHCF}~\cite{ji2020dual}: This model adopts dual-channel hypergraph neural networks for both users and items in collaborative filtering. 
\end{itemize}

\noindent \textbf{Recommenders enhanced by Self-Supervised Learning .}
\begin{itemize}[leftmargin=*]
\item \textbf{MHCN}~\cite{yu2021self}: This model maximizes the mutual information between node embeddings and global readout representations, to regularize the representation learning for interaction graph.
\item \textbf{SLRec}~\cite{yao2021self}: This approach employs the contrastive learning between the node features as regularization terms to enhance the existing recommender systems.
\item \textbf{SGL}~\cite{wu2021self}: This model conducts data augmentation through random walk and feature dropout to generate multiple views. It enhances LightGCN with self-supervised contrastive learning.
\end{itemize}

\subsubsection{\bf Implementation Details}
We implement our \model\ using TensorFlow and use Adam as the optimizer for model training with the learning rate of $1e^{-3}$ and $0.96$ epoch decay ratio. The models are configured with $32$ embedding dimension size, and the number of graph neural layers is searched from \{1,2,3\}. The weights $\lambda_1, \lambda_2$ for regularization terms are selected from $\{a\times 10 ^ {-x}: a\in\{1, 3\}, x\in\{2,3,4,5\}\}$. The batch size is selected from $\{32, 64, 128, 256, 512\}$. The rate for dropout is tuned from $\{0.25, 0.5, 0.75\}$. For our model, the number of hyperedges is set as $128$ by default. Detailed hyperparameter settings can be found in our released source code.

\subsection{Overall Performance Comparison (RQ1)}
\begin{table*}[t]
\vspace{-0.1in}
\caption{Performance comparison on Yelp, MovieLens, Amazon datasets in terms of \textit{Recall} and \textit{NDCG}.}
\vspace{-0.15in}
\centering
\footnotesize
\setlength{\tabcolsep}{1mm}
\begin{tabular}{|c|c|c|c|c|c|c|c|c|c|c|c|c|c|c|c|c|c|c|}
\hline
Data & Metric & BiasMF & NCF & AutoR & GCMC & PinSage & NGCF & STGCN & LightGCN & GCCF & DGCF & HyRec & DHCF & MHCN & SLRec & SGL & \emph{\model} & p-val.\\
\hline
\multirow{4}{*}{Yelp}
&Recall@20 & 0.0190 & 0.0252 & 0.0259 &  0.0266 & 0.0345 & 0.0294 & 0.0309 & 0.0482 & 0.0462 & 0.0466 & 0.0472 & 0.0449 & 0.0503 & 0.0476 & 0.0526 & \textbf{0.0651} & $9.3e^{-7}$\\
&NDCG@20 & 0.0161 & 0.0202 & 0.0210 & 0.0251 & 0.0288 & 0.0243 & 0.0262 & 0.0409 & 0.0398 & 0.0395 & 0.0395 & 0.0381 & 0.0424 & 0.0398 & 0.0444 & \textbf{0.0546} & $9.1e^{-8}$ \\
\cline{2-19}
&Recall@40 & 0.0371 & 0.0487 & 0.0504 & 0.0585 & 0.0599 & 0.0522 & 0.0504 & 0.0803 & 0.0760 & 0.0774 & 0.0791 & 0.0751 & 0.0826 & 0.0821 & 0.0869 & \textbf{0.1091} & $4.1e^{-7}$\\
&NDCG@40 & 0.0227 & 0.0289 & 0.0301 & 0.0373 & 0.0385 & 0.0330 & 0.0332 & 0.0527 & 0.0508 & 0.0511 & 0.0522 & 0.0493 & 0.0544 & 0.0541 & 0.0571 & \textbf{0.0709} & $2.2e^{-7}$ \\
\hline

\multirow{4}{*}{Gowalla}
&Recall@20 & 0.0196 & 0.0171 & 0.0239 & 0.0301 & 0.0576 & 0.0552 & 0.0369 & 0.0985 & 0.0951 & 0.0944 & 0.0901 & 0.0931 & 0.0955 & 0.0925 & 0.1030 & \textbf{0.1232} & $5.3e^{-7}$\\
&NDCG@20 & 0.0105 & 0.0106 & 0.0132 & 0.0181 & 0.0373 & 0.0298 & 0.0217 & 0.0593 & 0.0535 & 0.0522 & 0.0498 & 0.0505 & 0.0574 & 0.0581 & 0.0623 & \textbf{0.0731} & $6.3e^{-7}$\\
\cline{2-19}
&Recall@40 & 0.0346 & 0.0216 & 0.0343 & 0.0427 & 0.0892 & 0.0810 & 0.0542 & 0.1431 & 0.1392 & 0.1401 & 0.1306 & 0.1356 & 0.1393 & 0.1305 & 0.1500 & \textbf{0.1804} & $1.5e^{-7}$\\
&NDCG@40 & 0.0145 & 0.0118 & 0.0160 & 0.0212 & 0.0417 & 0.0367 & 0.0262 & 0.0710 & 0.0684 & 0.0671 & 0.0669 & 0.0660 & 0.0689 & 0.0680 & 0.0746 & \textbf{0.0881} & $3.2e^{-7}$\\
\hline

\multirow{4}{*}{Tmall}
&Recall@20 & 0.0103 & 0.0082 & 0.0103 & 0.0103 & 0.0202 & 0.0180 & 0.0146 & 0.0225 & 0.0209 & 0.0235 & 0.0233 & 0.0156 & 0.0203 & 0.0191 & 0.0268 & \textbf{0.0387} & $4.3e^{-9}$\\
&NDCG@20 & 0.0072 & 0.0059 & 0.0072 & 0.0072 & 0.0136 & 0.0123 & 0.0105 & 0.0154 & 0.0141 & 0.0163 & 0.0160 & 0.0108 & 0.0139 & 0.0133 & 0.0183 & \textbf{0.0262} & $4.9e^{-9}$\\
\cline{2-19}
&Recall@40 & 0.0170 & 0.0140 & 0.0174 & 0.0159 & 0.0345 & 0.0310 & 0.0245 & 0.0378 & 0.0356 & 0.0394 & 0.0350 & 0.0261 & 0.0340 & 0.0301 & 0.0446 & \textbf{0.0645} & $4.0e^{-9}$\\
&NDCG@40 & 0.0095 & 0.0079 & 0.0097 & 0.0086 & 0.0186 & 0.0168 & 0.0140 & 0.0208 & 0.0196 & 0.0218 & 0.0199 & 0.0145 & 0.0188 & 0.0171 & 0.0246 & \textbf{0.0352} & $3.5e^{-9}$\\
\hline
\end{tabular}
\label{tab:overall_performance}
\end{table*}

In this section, we validate the effectiveness of our \model\ framework by conducting the overall performance evaluation on the three datasets and comparing \model\ with various baselines. We also re-train \model\ and the best-performed baseline (\ie~SGL) for 10 times to compute p-values. The results are presented in Table~\ref{tab:overall_performance}.
\begin{itemize}[leftmargin=*]
    \item \textbf{Performance Superiority of \model}. As shown in the results, \model\ achieves best performance compared to the baselines under both top-\textit{20} and top-\textit{40} settings. The t-tests also validate the significance of performance improvements. We attribute the superiority to: i) Based on the hypergraph transformer, \model\ not only realizes global message passing among semantically-relevent users/items, but also refines the hypergraph structure using the multi-head attention. ii) The global-to-local self-augmented learning distills knowledge from the high-level hypergraph transformers to regularize the topology-aware embedding learning, and thus alleviate the data noise issue.
    
    \item \textbf{Effectiveness of Hypergraph Architecture}. Among the state-of-the-art baselines, approaches that based on hypergraph neural networks (HGNN) (\ie, HyRec and DHCF) outperforms most of the GNN-based baselines (\eg, GCMC, PinSage, NGCF, STGCN). This sheds lights on the insufficiency of conventional GNNs in capturing high-order and global graph connectivity. Meanwhile, our \model\ is configured with transformer-like hypergraph structure learning which further excavates the potential of HGNN in global relation learning. In addition, most existing hypergraph-based models utilize user or item nodes as hyperedges, while our \model\ adopts latent hyperedges which not only enables automatic graph dependency modeling, but also avoids pre-calculating the large-scale high-order relation matrix.
    
    \item \textbf{Effectiveness of Self-Augmented Learning}. From the evaluation results, we can observe that self-supervised learning obviously improves existing CF frameworks (\eg, MHCN, SLRec, SGL). The improvements can be attributed to incorporating the augmented learning task, which provides the beneficial regularization on the parameter learning based on the input data itself. Specifically, MHCN regularizes the node embeddings according to a read-out global information of the holistic graph. This approach may be too strict for large graphs containing many local sub-graphs with their own characteristics. Meanwhile, SLRec and SGL adopt stochastic data augmentation to construct multiple data views, and conduct contrastive learning to capture the invariant feature from the corrupted views. In comparison to the above methods, the self-augmentation in our \model\ has mainly two merits: i) \model\ adopts meta networks to generate global-structure-aware mapping functions for domain adaption, which adaptively alleviates the gap between local and global feature spaces. ii) Our self-supervised approach does not depend on random masking, which may drop important information to hinder representation learning. Instead, \model\ self-augment the model training by transferring knowledge from the high-level hypergraph embeddings to the low-level topology-aware embedding. The superior performance of \model\ compared to the baseline self-supervised approaches validates the effectiveness of our new design of self-supervised learning paradigm.
\end{itemize}

\vspace{-0.1in}
\subsection{Model Ablation Test (RQ2)}
To validate the effectiveness of the proposed modules, we individually remove the applied techniques in the three major parts of \model\ (\ie, the local graph structure capturing, the global relation learning, and the local-global self-augmented learning). The variants are re-trained for test on the three datasets. Both prominent components (\eg, the entire hypergraph transformer) and small modules (\eg, the deep hyperedge feature extraction) of \model\ are ablated. The results can be seen in Table~\ref{tab:module_ablation}. We have the following major conclusions:
\begin{itemize}[leftmargin=*]
    \item Removing either the graph topology-aware embedding module or the hypergraph transformer (\ie, \textit{-Pos} and \textit{-Hyper}) severely damage the performance of \model\ in all cases. This result suggests the necessity of local and global relation learning, and validates the effectiveness of our GCN-based topology-aware embedding and hypergraph transformer networks.
    
    \item The variant without self-augmented learning (\ie~\textit{-SAL}) yields obvious performance degradation in all cases, which validates the positive effect of our augmented global-to-local knowledge transferring. The effect of our meta-network-based domain adaption can also be observed in the variant \textit{-Meta}.
    
    \item We also ablate the components in our hypergraph neural network. Specifically, we substitute the hypergraph transformer with independent node-hypergraph matrices (\textit{-Trans}), and we remove the deep hyperedge feature extraction to keep only one layer of hyperedges (\textit{-DeepH}). Additionally, we remove the high-order hypergraph iterations (\textit{-HighH}). From the results we can conclude that: i) Though using much less parameters, the transformer-like hypergraph attention works much better than learning hypergraph-based user/item dependencies. ii) The deep hyperedge layers indeed make contribution to the global relation learning through non-linear feature transformation. iii) Though our hypergraph transformer could connect any users/items using learnable hyperedges, high-order iterations still improve the model performance through the iterative hypergraph propagation.
\end{itemize}

\begin{table}[t]
    \caption{Ablation study on key components of \model.}
    \vspace{-0.15in}
    \centering
    \footnotesize
    \setlength{\tabcolsep}{1.2mm}
    \begin{tabular}{c|c|cc|cc|cc}
        \hline
        \multirow{2}{*}{Category} & Data & \multicolumn{2}{c|}{Yelp} & \multicolumn{2}{c|}{Gowalla} & \multicolumn{2}{c}{Tmall}\\
        \cline{2-8}
        & Variants & Recall & NDCG & Recall & NDCG & Recall & NDCG\\
        \hline
        \hline
        Local & -Pos & 0.0423 & 0.0352 & 0.0816 & 0.0487 & 0.0218 & 0.0247\\
        \hline
        \multirow{4}{*}{Global} & -Trans & 0.0603 & 0.0504 & 0.0999 & 0.0608 & 0.0321 & 0.0206\\
        &-DeepH & 0.0645 & 0.0540 & 0.1089 & 0.0634 &  0.0347 & 0.0234\\
        &-HighH & 0.0598 & 0.0497 & 0.1091 & 0.0646 & 0.0336 & 0.0227\\
        &-Hyper & 0.0401 & 0.0346 & 0.0879 & 0.0531 & 0.0209 & 0.0144\\
        \hline
        \multirow{2}{*}{SAL} & -Meta & 0.0615 & 0.0526 & 0.1108 & 0.0717 & 0.0375 & 0.0255\\
        &-SAL & 0.0602 & 0.0519 & 0.1099 & 0.0699 & 0.0363 & 0.0251\\
        \hline
        \multicolumn{2}{c|}{\emph{\model}} & 0.0651 & 0.0546 & 0.1232 & 0.0731 & 0.0387 & 0.0262\\
        \hline
    \end{tabular}
    \vspace{-0.15in}
    \label{tab:module_ablation}
\end{table}

\subsection{Model Robustness Test (RQ3)}
\subsubsection{\bf Performance \textit{w.r.t.} Data Noise Degree}
\begin{figure}[t]
    \centering
    \subfigure[Yelp data]{
        \includegraphics[width=0.47\columnwidth]{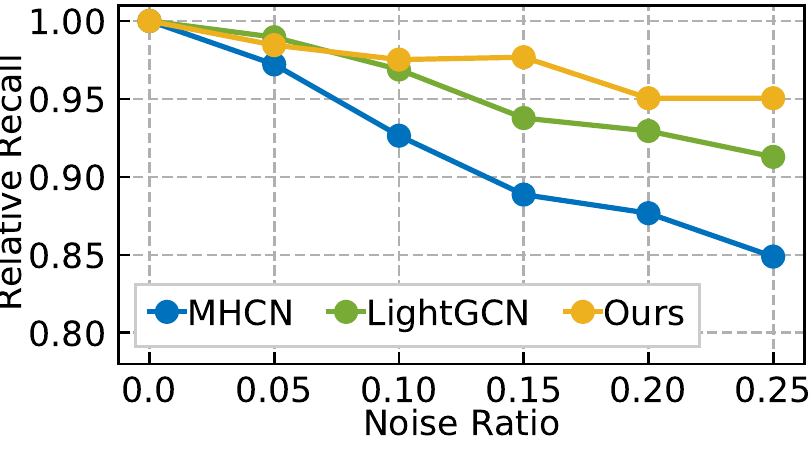}\ 
        \includegraphics[width=0.47\columnwidth]{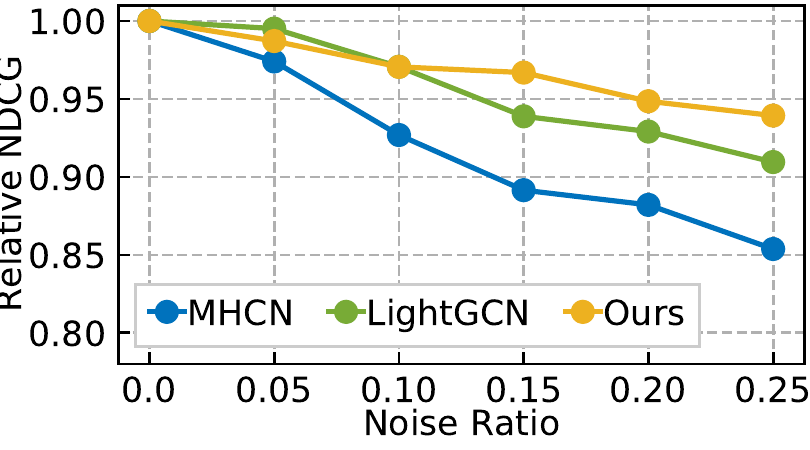}
        \vspace{-0.15in}
    }
    \subfigure[Gowalla data]{
        \includegraphics[width=0.47\columnwidth]{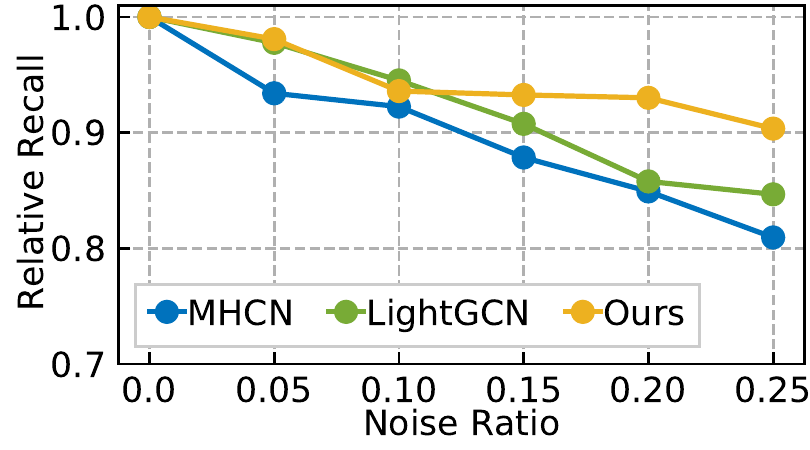}\ 
        \includegraphics[width=0.47\columnwidth]{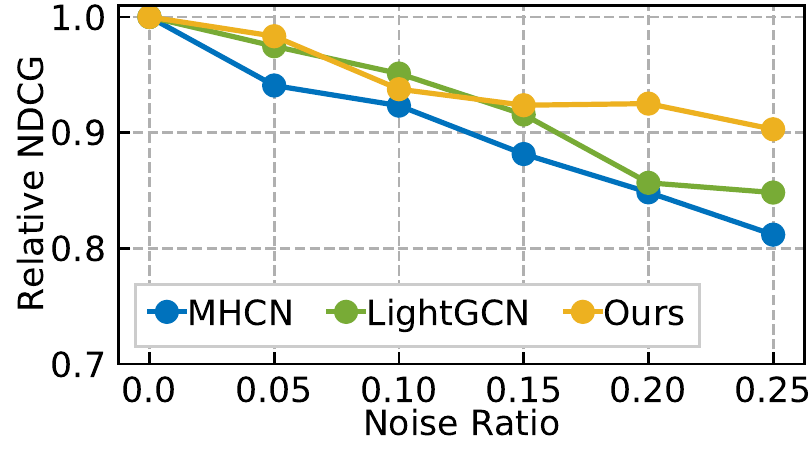}
    }
    \subfigure[Tmall data]{
        \includegraphics[width=0.47\columnwidth]{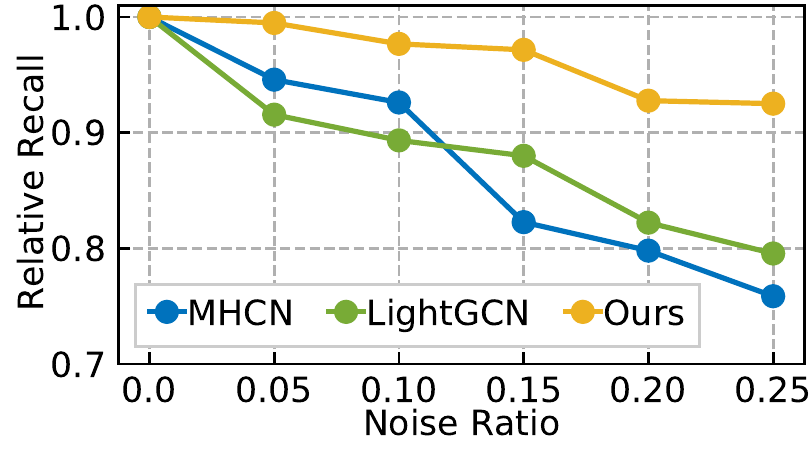}\ 
        \includegraphics[width=0.47\columnwidth]{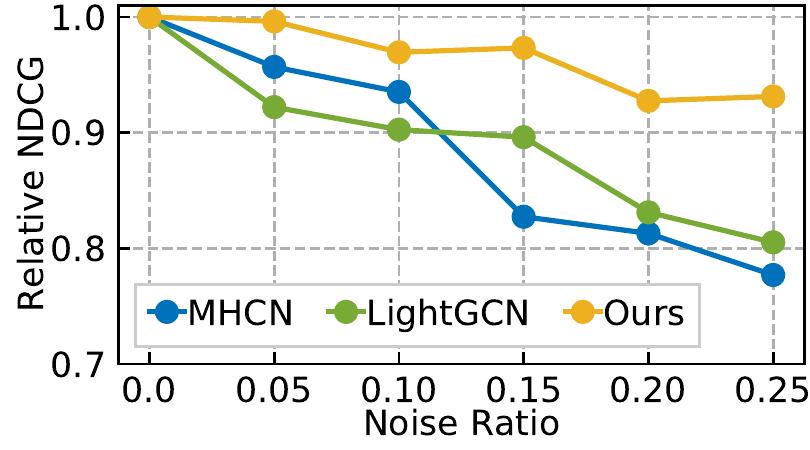}
    }
    \vspace{-0.15in}
    \caption{Relative performance degradation \wrt\ noise ratio.}
    \vspace{-0.2in}
    \label{fig:noise}
\end{figure}

In this section, we first investigate the robustness of \model\ against the data noise. To evaluate the influence of noise degrees on model performance, we randomly substitute different percentage of real edges with randomly-generated fake edges, and re-train the model using the corrupted graphs as input. Concretely 5\%, 10\%, 15\%, 20\%, 25\% of the edges are replaced with noisy signals in our experiments. We compare \model\ with MHCN and LightGCN, which are recent recommenders based on HGNN and GNN, respectively. To better study the effect of noise on performance degradation, we evaluate the relative performance compared to the performance on original data. The results are shown in Fig~\ref{fig:noise}. We can observe that our method presents smaller performance degradation in most cases compared to the baselines. We ascribe this observation to two reasons: i) The global relation learning and information propagation by our hypergraph transformer alleviate the noise effect caused by the raw observed user-item interactions. ii) The self-augmented learning task distills knowledge from the refined hypergraph embeddings, so as to refine the graph-based embeddings. In addition, we can observe that the relative performance degradation on the Gowalla data is more obvious compared with other two datasets. 
This is because the noisy data has larger influence for the performance on the sparsest Gowalla dataset.

\subsubsection{\bf Performance \textit{w.r.t.} Data Sparsity}
We further study the influence of data sparsity from both user and item side on model performance. We compare our \model\ with two representative baselines LightGCN and SGL.
Multiple user and item groups are constructed in terms of their number of interactions in the training set.
For example, the first group in the user-side experiments contains users interacting with 15-20 items, and the first group in the item-side experiment contains items interacting with 0-8 users.

\begin{figure}[t]
    \centering
    \subfigure[Performance \textit{w.r.t.} item interaction numbers]{
        \includegraphics[width=0.47\columnwidth]{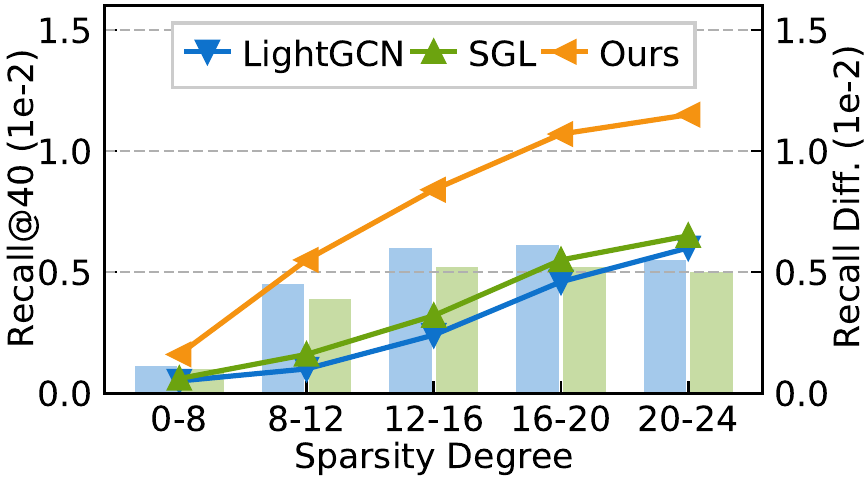}\ 
        \includegraphics[width=0.47\columnwidth]{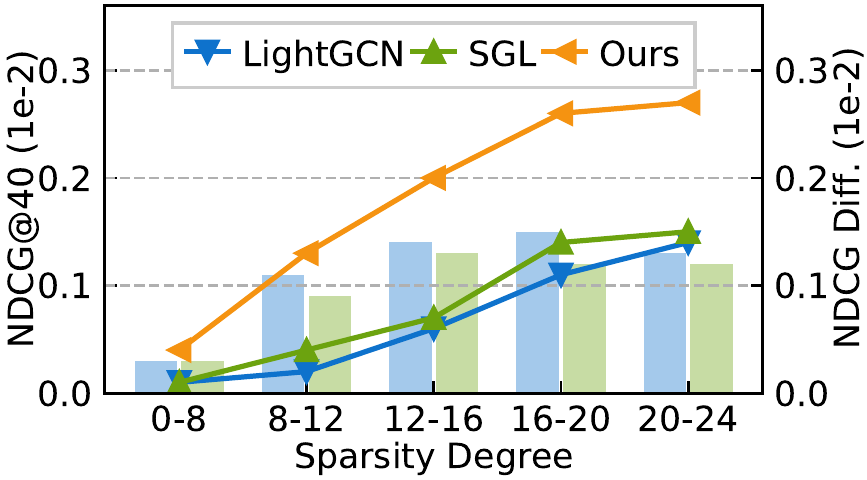}
        \vspace{-0.15in}
    }
    \subfigure[Performance \textit{w.r.t.} user interaction numbers]{
        \includegraphics[width=0.47\columnwidth]{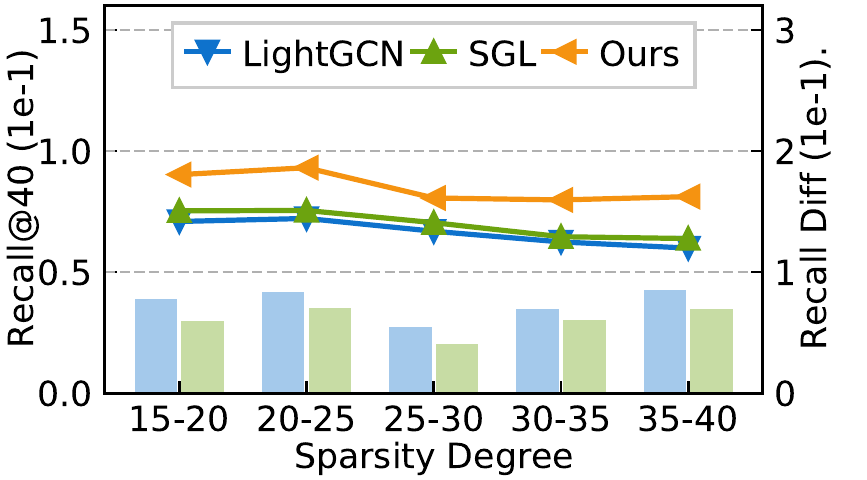}\ 
        \includegraphics[width=0.47\columnwidth]{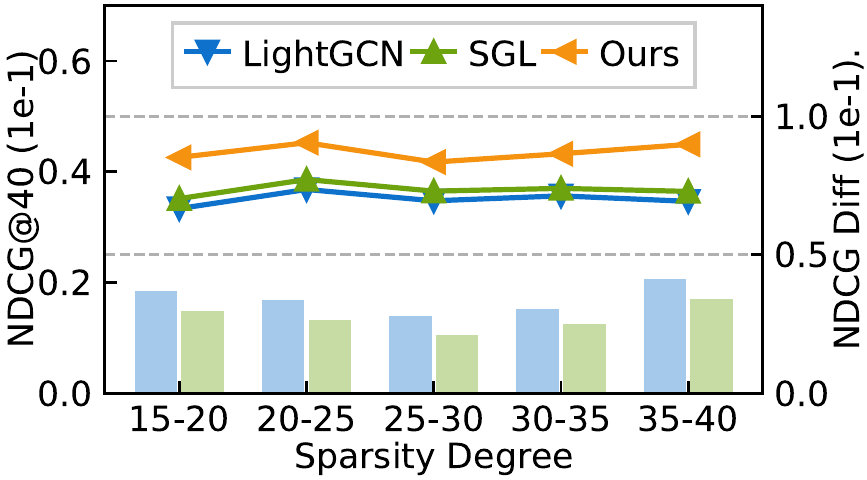}
    }
    \vspace{-0.15in}
    \caption{Performance \textit{w.r.t.} different data sparsity degrees on Gowalla data. Lines present Recall@40 and NDCG@40 values, and bars shows performance differences between baselines and our \model\ with corresponding colors.}
    \vspace{-0.15in}
    \label{fig:sparse}
\end{figure}

In Fig~\ref{fig:sparse}, we present both the recommendation accuracy and performance difference between our \model\ and compared methods.
From the results, we have the following observations: i) The superior performance of \model\ is consistent on datasets with different sparsity degrees, which validates the robustness of \model\ in handling sparse data for both users and items. ii) The sparsity of item interaction vectors has obviously larger influence on model performance for all the methods. This indicates that the collaborative pattern of items are more difficult to model compared to users, such that more neighbors usually result in better representations. iii) In the item-side experiments, the performance gap on the middle sub-datasets is larger compared to the gap on the densest sub-dataset. This suggests the better anti-sparsity capability of \model\ for effectively transferring knowledge among dense samples and sparse samples with our proposed hypergraph transformer.

\subsection{Case Study (RQ4)}
\begin{figure}
    \centering
    \includegraphics[width=0.95\columnwidth]{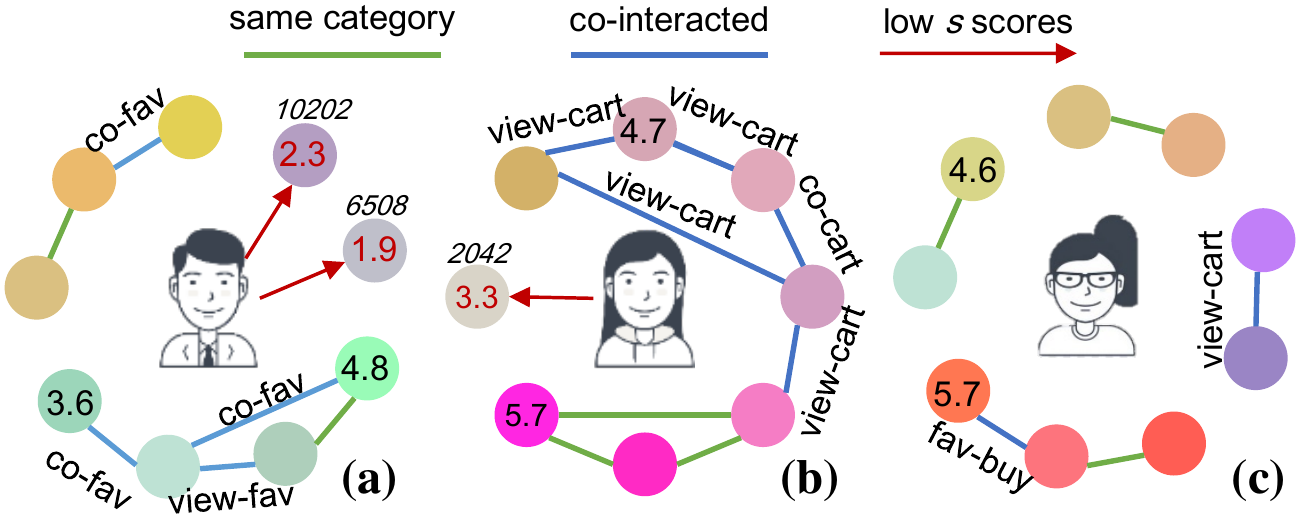}
    \caption{Case study on inferring implicit item-wise relations and discriminating potential noise edges. Circles denote items interacted by the centric users, and their learned embeddings are visualized with colors. Implicit item-wise relations not utilized during model training are presented by green and blue lines. The type of co-interactions are also labeled (\eg, \textit{view-cart} denotes viewed and added-to-cart by same users). Also, the inferred solidity scores $s$ are shown on the circles, where red values are anomalously low scores indicating noisy edges.}
    \label{fig:case_study}
    \vspace{-0.1in}
\end{figure}

In this section, we analyze the concrete data instances to investigate the effect of our hypergraph transformer with self-augmentation from two aspects: i) Is the hypergraph-based dependency modeling in \model\ capable of learning useful node-wise relations, especially the implicit relations unknown to the training process? ii) Is the self-augmented learning with meta networks in \model\ able to differentiate noisy edges in the training data? To this end, we select three users with fair number of interactions from Tmall dataset. The interacted items are visualized as colored circles representing their trained embeddings (refer to the supplementary material for details about the visualization algorithm). The results are shown in Fig~\ref{fig:case_study}. For the above questions, we have the following observations:

\begin{itemize}[leftmargin=*]
    \item \textbf{Implicit relation learning}. Even if the items are interacted by same users, their learned embeddings are usually divided into multiple groups with different colors. This may relate to users' multiple interests. To study the differences between the item groups, we present additional item-wise relations that are not utilized in the training process. Specifically, we connect items belonging to same categories, and items co-interacted by same users. Note that only view data is used in model training, so interactions in other behaviors are unknown to the trained model. It is clear that there exist dense implicit correlations among same-colored items (\eg, the green items of user (a), the purple items of user (b), and the orange items of user (c)). Meanwhile, there are much less implicit relations between items of different colors. This results shows the capability of \model\ in identifying useful implicit relations, which we ascribe to the global structure learning of our hypergraph transformer.
    
    \item \textbf{Noise discrimination}. Furthermore, we show the solidity scores $s$ estimated from our self-augmented learning, for the user-item relations in Fig~\ref{fig:case_study}. We also show the normalized values of some notable edges in the corresponding circles (\eg, edges of item 10202 and 6508 are labeled with 2.3 and 1.9). The red values are anomalously low, which may indicates noise. The black values are the lowest and highest solidity scores for edges except the anomalous ones. By analyzing user (a), we can regard the yellow and green items as two interests of user (a) as they are correlated in terms of their learned embeddings. In contrast, item 6508 and 10202 have few relations to other interacted items of user (a), which may not reflect the real interactive patterns of this user. Thus, the model may consider this two edges as noisy interactions. Similar cases can be found for user (b), where item 2042 has few connections to the other items and show difference with the embedding color. It is labeled with low $s$ scores and considered as noise by \model. The results show the effective noise discrimination ability of the self-augmented learning in \model, which recalibrate the topology-aware embedding using global information encoded from hypergraph transformer.
\end{itemize}

\section{Conclusion}
\label{sec:conclusion}

In this work, we explore the self-supervised recommender systems with an effective hypergraph transformer network. We propose a new recommendation framework \model, which seeks better user-item interaction modeling with self-augmented supervision signals. Our \model\ model improves the robustness of graph-based recommender systems against noise perturbation. In our experiments, we achieved better recommendation results on real-world datasets. Our future work would like to extend our \model\ to explore the disentangled user intents with diverse user-item relations for encoding multi-dimensional user preferences.




\section*{Acknowledgments}
This research work is supported by the research grants from the Department of Computer Science \& Musketeers Foundation Institute of Data Science at the University of Hong Kong.

\bibliographystyle{ACM-Reference-Format}
\balance
\bibliography{sigproc.bib}

\clearpage
\section{Supplemental Material}
\label{sec:appendix}

%

In the supplementary material, we first show the learning process of \model\ with pseudocode summarized in Algorithm~\ref{alg:learn_alg}. Then, we investigate the influences of different hyperparameter settings, and discuss the impact of three key hyperparameters. Finally, we describe the details of our vector visualization algorithm used in the case study experiments.

\subsection{Learning process of \model}
\vspace{-0.1in}

\begin{algorithm}[h]
	\caption{Learning Process of \model\ Framework}
	\label{alg:learn_alg}
	\LinesNumbered
	\KwIn{user-item interaction graph $\mathcal{G}$, number of graph layers $L$, number of edges to sample $R, R'$, maximum epoch number $E$, learning rate $\eta$}
	\KwOut{trained parameters in $\mathbf{\Theta}$}
	Initialize all parameters in $\mathbf{\Theta}$\\
    \For{$e=1$ to $E$}{
        Draw a mini-batch $\textbf{U}$ from all users $\{1,2,...,I\}$\\
        Calculate the graph topology-aware embeddings $\bar{\textbf{E}}$\\
        Generate input embeddings $\tilde{\textbf{E}}_0$ for hypergraph transformer\\
        \For{$l=1$ to $L$} {
            Conduct node-to-hyperedge propagation to obtain $\tilde{\textbf{Z}}^{(u)}, \tilde{\textbf{Z}}^{(v)}$ for both users and items\\
            Conduct hierarchical hyperedge feature transformation for $\hat{\textbf{Z}}^{(u)}, \hat{\textbf{Z}}^{(v)}$\\
            Propagate information from hyperedges back to user/item nodes to obtain $\tilde{\textbf{E}}^{(u)}_l,\tilde{\textbf{E}}^{(v)}_l$\\
        }
        Aggregate the iteratively propagated embeddings to get $\hat{\textbf{E}}$\\
        
        Sample $R$ edge pairs for self-augmented learning\\
        Acquire the user/item transformation function $\phi^{(u)}$ and $\phi^{(v)}$ with the meta network\\
        Conduct user/item embedding transformations using $\phi(\cdot)$ to get $\mathbf{\Gamma}^{(u)}, \mathbf{\Gamma}^{(v)}$\\
        Calculate the solidity score $s$ for the $R$ edge pairs\\
        Calculate the solidity predictions $\hat{s}$ for the $R$ edge pairs\\
        Compute loss $\mathcal{L}_\text{sa}$ for self-augmented learning according to Eq~\ref{eq:sa_loss}\\
        
        Sample $R'$ edge pairs for the main task\\
        Calculate the pair-wise marginal loss $\mathcal{L}$ according to Eq~\ref{eq:loss}\\
        
        \For{each parameter $\theta \in\mathbf{\Theta}$}{
            $\theta=\theta-\eta\cdot\partial\mathcal{L}/\partial\theta$
        }
    }
    \Return all parameters $\mathbf{\Theta}$
\end{algorithm}

\subsection{Hyperparameter Investigation}
We study the effect of three important hyperparameters, \ie, the hidden dimensionality $d$, the number of latent hyperedges $K$, and the number of graph iterations $L$. To present more results, we calculate the relative performance decrease in terms of evaluation metrics, compared to the best performance under default settings. The results are shown in Fig~\ref{fig:hyperparam}, our observations are shown below:
\begin{itemize}[leftmargin=*]
    \item The latent embedding dimension size largely determines the representation ability of the proposed \model\ model. Small $d$ greatly limits the efficacy of \model, by 15\%-35\% performance decrease. However, greater $d$ does not always yield obvious improvements. As shown by results when $d=68$ on Yelp data, the performance increases marginally due to the over-fitting effect.
    
    \item The curve of performance w.r.t. hyperedge number $K$ typically follows the under- to over-fitting pattern. However, it is interesting that $K$ has significantly less influence compared to $d$ (at most $-6\%$ and $-35\%$, respectively). This is because the hyperedge-node connections in \model\ are calculated in a $d$-dimensional space, which reduces the amount of independent parameters related to $K$ to $O(K\times d)$. So $K$ has much smaller impact on model capacity compared to $d$, which relates to $O((I+J)\times d)$ parameters.
    
    \item For the number of graph iterations $L$, smaller $L$ hinders nodes from aggregating high-order neighboring information. When $L=0$, graph neural networks degrades significantly. Meanwhile, by stacking more graph layers may cause over-smoothing issue, which yields indistinguishable node embeddings.
\end{itemize}
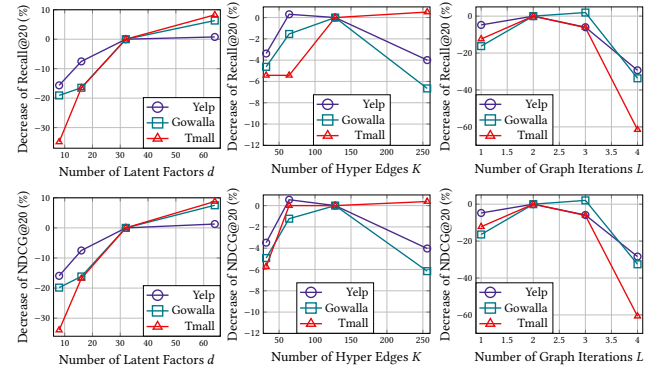
\begin{figure}[h]
    \centering
    \begin{adjustbox}{max width=1.0\linewidth}
    \begin{tikzpicture}
\begin{axis}[
    xlabel={Number of Latent Factors $d$},
    ylabel={Decrease of Recall@20 (\%)},
    xmin=6,xmax=66,
    ymin=-37,ymax=10,
    legend columns=1,
    legend cell align=right,
    grid=both,
    every axis plot/.append style={ultra thick},
    every tick label/.append style={scale=1.3},
    label style={scale=1.8},
    legend style={
        nodes={scale=1.5, transform shape},
        legend image post style={scale=1.5},
        },
    legend style={at={(1,0)},anchor=south east},
    every axis plot post/.append style={
        every mark/.append style={scale=2}
    }
]
\addplot[color={rgb:red,133;green,76;blue,255}, mark=o, mark options={solid}]
table[x=para, y=yelp_hr] {latFactor.txt};
\addplot[color={rgb:red,0;green,157;blue,178}, mark=square, mark options={solid}]
table[x=para, y=gowalla_hr] {latFactor.txt};
\addplot[color={rgb:red,245;green,9;blue,11}, mark=triangle, mark options={solid}]
table[x=para, y=tmall_hr] {latFactor.txt};
\legend{\large Yelp, \large Gowalla, \large Tmall};
\end{axis}
\end{tikzpicture}

\begin{tikzpicture}
\begin{axis}[
    xlabel={Number of Hyper Edges $K$},
    ylabel={Decrease of Recall@20 (\%)},
    xmin=28,xmax=260,
    ymin=-12,ymax=1,
    legend columns=1,
    legend cell align=right,
    grid=both,
    every axis plot/.append style={ultra thick},
    every tick label/.append style={scale=1.3},
    label style={scale=1.8},
    legend style={
        nodes={scale=1.5, transform shape},
        legend image post style={scale=1.5},
        },
    legend style={at={(0.3,0)},anchor=south west},
    every axis plot post/.append style={
        every mark/.append style={scale=2}
    }
]
\addplot[color={rgb:red,133;green,76;blue,255}, mark=o, mark options={solid}]
table[x=para, y=yelp_hr] {hyperNum.txt};
\addplot[color={rgb:red,0;green,157;blue,178}, mark=square, mark options={solid}]
table[x=para, y=gowalla_hr] {hyperNum.txt};
\addplot[color={rgb:red,245;green,9;blue,11}, mark=triangle, mark options={solid}]
table[x=para, y=tmall_hr] {hyperNum.txt};
\legend{\large Yelp, \large Gowalla, \large Tmall};
\end{axis}
\end{tikzpicture}

\begin{tikzpicture}
\begin{axis}[
    xlabel={Number of Graph Iterations $L$},
    ylabel={Decrease of Recall@20 (\%)},
    xmin=0.9,xmax=4.1,
    ymin=-70,ymax=5,
    legend columns=1,
    legend cell align=right,
    grid=both,
    every axis plot/.append style={ultra thick},
    every tick label/.append style={scale=1.3},
    label style={scale=1.8},
    legend style={
        nodes={scale=1.5, transform shape},
        legend image post style={scale=1.5},
        },
    legend style={at={(0,0)},anchor=south west},
    every axis plot post/.append style={
        every mark/.append style={scale=2}
    }
]
\addplot[color={rgb:red,133;green,76;blue,255}, mark=o, mark options={solid}]
table[x=para, y=yelp_hr] {gnnlayer.txt};
\addplot[color={rgb:red,0;green,157;blue,178}, mark=square, mark options={solid}]
table[x=para, y=gowalla_hr] {gnnlayer.txt};
\addplot[color={rgb:red,245;green,9;blue,11}, mark=triangle, mark options={solid}]
table[x=para, y=tmall_hr] {gnnlayer.txt};
\legend{\large Yelp, \large Gowalla, \large Tmall};
\end{axis}
\end{tikzpicture}
    \end{adjustbox}
    \begin{adjustbox}{max width=1.0\linewidth}
    \begin{tikzpicture}
\begin{axis}[
    xlabel={Number of Latent Factors $d$},
    ylabel={Decrease of NDCG@20 (\%)},
    xmin=6,xmax=66,
    ymin=-36,ymax=10,
    legend columns=1,
    legend cell align=right,
    grid=both,
    every axis plot/.append style={ultra thick},
    every tick label/.append style={scale=1.3},
    label style={scale=1.8},
    legend style={
        nodes={scale=1.5, transform shape},
        legend image post style={scale=1.5},
        },
    legend style={at={(1,0)},anchor=south east},
    every axis plot post/.append style={
        every mark/.append style={scale=2}
    }
]
\addplot[color={rgb:red,133;green,76;blue,255}, mark=o, mark options={solid}]
table[x=para, y=yelp_ndcg] {latFactor.txt};
\addplot[color={rgb:red,0;green,157;blue,178}, mark=square, mark options={solid}]
table[x=para, y=gowalla_ndcg] {latFactor.txt};
\addplot[color={rgb:red,245;green,9;blue,11}, mark=triangle, mark options={solid}]
table[x=para, y=tmall_ndcg] {latFactor.txt};
\legend{\large Yelp, \large Gowalla, \large Tmall};
\end{axis}
\end{tikzpicture}

\begin{tikzpicture}
\begin{axis}[
    xlabel={Number of Hyper Edges $K$},
    ylabel={Decrease of NDCG@20 (\%)},
    xmin=28,xmax=260,
    ymin=-12,ymax=1,
    legend columns=1,
    legend cell align=right,
    grid=both,
    every axis plot/.append style={ultra thick},
    every tick label/.append style={scale=1.3},
    label style={scale=1.8},
    legend style={
        nodes={scale=1.5, transform shape},
        legend image post style={scale=1.5},
        },
    legend style={at={(0.2,0)},anchor=south west},
    every axis plot post/.append style={
        every mark/.append style={scale=2}
    }
]
\addplot[color={rgb:red,133;green,76;blue,255}, mark=o, mark options={solid}]
table[x=para, y=yelp_ndcg] {hyperNum.txt};
\addplot[color={rgb:red,0;green,157;blue,178}, mark=square, mark options={solid}]
table[x=para, y=gowalla_ndcg] {hyperNum.txt};
\addplot[color={rgb:red,245;green,9;blue,11}, mark=triangle, mark options={solid}]
table[x=para, y=tmall_ndcg] {hyperNum.txt};
\legend{\large Yelp, \large Gowalla, \large Tmall};
\end{axis}
\end{tikzpicture}

\begin{tikzpicture}
\begin{axis}[
    xlabel={Number of Graph Iterations $L$},
    ylabel={Decrease of NDCG@20 (\%)},
    xmin=0.9,xmax=4.1,
    ymin=-70,ymax=5,
    legend columns=1,
    legend cell align=right,
    grid=both,
    every axis plot/.append style={ultra thick},
    every tick label/.append style={scale=1.3},
    label style={scale=1.8},
    legend style={
        nodes={scale=1.5, transform shape},
        legend image post style={scale=1.5},
        },
    legend style={at={(0,0)},anchor=south west},
    every axis plot post/.append style={
        every mark/.append style={scale=2}
    }
]
\addplot[color={rgb:red,133;green,76;blue,255}, mark=o, mark options={solid}]
table[x=para, y=yelp_ndcg] {gnnlayer.txt};
\addplot[color={rgb:red,0;green,157;blue,178}, mark=square, mark options={solid}]
table[x=para, y=gowalla_ndcg] {gnnlayer.txt};
\addplot[color={rgb:red,245;green,9;blue,11}, mark=triangle, mark options={solid}]
table[x=para, y=tmall_ndcg] {gnnlayer.txt};
\legend{\large Yelp, \large Gowalla, \large Tmall};
\end{axis}
\end{tikzpicture}
    \end{adjustbox}
    \vspace{-0.2in}
    \caption{Hyperparameter study of the \model.}
    \vspace{-0.2in}
    \label{fig:hyperparam}
\end{figure}

\subsection{Vector Visualization Algorithm}
In our case study experiments, each item embeddings of $32$ dimensions is visualized with a color. This visualization process should preserve the learned item information in the embedding vectors. Meanwhile, to make the visualization results easy to understand, it would be better to pre-select several colors to use. Considering the above two requirements, we design a neural-network-based dimension reduction algorithm. Specifically, we train a multi-layer perceptron to map 32-dimensional item embeddings to 3-dimensional RGB values. The network is trained using two objective functions, corresponding to the forgoing two requirements. Firstly, the compressed 3-d vectors (colors) are fed into a classifier, to predict the original item ids. Through this self-discrimination task, the network is trained to preserve the original embedding information in the RGB vectors. Secondly, the network is trained with a regularizer that calculates the distance between each color vectors and the preferred colors. Using the two objectives, we can map embeddings into preferred colors while preserving the embedding information.

\end{document}